\documentclass{sigchi}

\CopyrightYear{2020}
\setcopyright{rightsretained}
\doi{https://doi.org/10.1145/3313831.3376409}
\isbn{978-1-4503-6708-0/20/04}
\conferenceinfo{CHI'20,}{April  25--30, 2020, Honolulu, HI, USA}

\usepackage{balance}       
\usepackage{graphics}      
\usepackage[T1]{fontenc}   
\usepackage{txfonts}
\usepackage{mathptmx}
\usepackage[pdflang={en-US},pdftex]{hyperref}
\usepackage{color}
\usepackage{booktabs}
\usepackage{textcomp}
\usepackage{siunitx}

\usepackage{subcaption}

\usepackage{microtype}        
\usepackage{ccicons}          

\usepackage{todonotes}

\def\plaintitle{Levitation Simulator: Prototyping Ultrasonic Levitation Interfaces in Virtual Reality}

\def\emptyauthor{}
\def\plainkeywords{Modeling; Simulation; Virtual Prototyping; Ultrasonic Levitation; VR.}

\makeatletter
\def\url@leostyle{%
  \@ifundefined{selectfont}{
    \def\UrlFont{\sf}
  }{
    \def\UrlFont{\small\bf\ttfamily}
  }}
\makeatother
\def\pprw{8.5in}
\def\pprh{11in}

\definecolor{linkColor}{RGB}{6,125,233}
\hypersetup{%
  pdftitle={\plaintitle},
  pdfauthor={\emptyauthor},
  pdfkeywords={\plainkeywords},
  pdfdisplaydoctitle=true, 
  bookmarksnumbered,
  pdfstartview={FitH},
  colorlinks,
  citecolor=black,
  filecolor=black,
  linkcolor=black,
  urlcolor=linkColor,
  breaklinks=true,
  hypertexnames=false
}

\begin{document}

\title{\plaintitle}

\numberofauthors{1}
\author{%
  \alignauthor{Viktorija Paneva \hspace{2cm} Myroslav Bachynskyi \hspace{2cm} J{\"o}rg M{\"u}ller\\
\affaddr{University of Bayreuth, Germany\\
\{viktorija.paneva, myroslav.bachynskyi, joerg.mueller\}@uni-bayreuth.de}}
}

\teaser{ 
\centering 
\includegraphics[width=\textwidth, height=5cm]{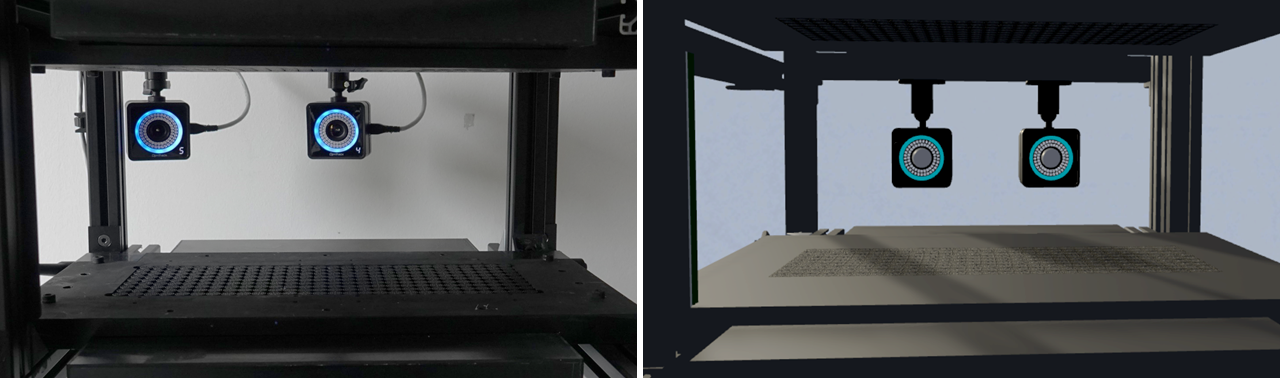}
\caption{The {\em Levitation Simulator} allows to simulate interaction with a levitation interface in VR. 
Two user studies, comparing the  real prototype (left) with the VR simulator (right), show that the VR simulation provides a good approximation of the interaction with the levitating particle in the real prototype. 
We share our {\em Levitation Simulator} as Open Source (www.ai8.uni-bayreuth.de/en/projects/Levisim), thereby democratizing levitation research and facilitating the design of applications for levitation interfaces, without the need for a levitation apparatus.} 
\label{fig:teaser} } 

\maketitle

\begin{abstract}
  We present the {\em Levitation Simulator}, 
  a system that enables researchers and designers to iteratively develop and prototype levitation interface ideas in Virtual Reality.
  This includes user tests and formal experiments.
  We derive a model of the movement of a levitating particle in such an interface.
  Based on this, we develop an interactive simulation of the levitation interface in VR, which exhibits the dynamical properties of the real interface.
 The results of a Fitts’ Law pointing study show that the Levitation Simulator enables performance, comparable to the real prototype. 
 We developed the first two interactive games, dedicated for levitation interfaces: LeviShooter and BeadBounce, in the Levitation Simulator, and then implemented them on the real interface. 
 Our results indicate that participants experienced similar levels of user engagement when playing the games, in the two environments.
     We share our {\em Levitation Simulator} as Open Source, thereby democratizing levitation research, without the need for a levitation apparatus. 
  
\end{abstract}




\section{Introduction}
Ultrasonic Levitation Interfaces promise a future in which the computer can control the existence of matter in our environment, truly merging real and virtual worlds.
This is similar to approaches such as Programmable Matter \cite{goldstein} and Radical Atoms \cite{ishii}.
However, with Ultrasonic Levition, power supply, actuation, and computation are placed in the environment making the individual atoms simpler and cheaper. 

Despite the potential of such interfaces and the accumulated body of research~\cite{hirayama, leviprops, marzonature, pointandshake, pixiedust, bachynskyi18}, not many applications for levitation interfaces have been developed thus far.
Among the few examples are a two-platform jump game \cite{JOLED}, and augmented static objects (e.g. volcano) \cite{EnhancingPhysicalObjects}.
The main reason that so few applications exist is that it is difficult for designers, artists, game developers and even researchers to work with ultrasonic levitation interfaces. 

Building and maintaining an interactive levitation interface, requires time, resources and very specific technical expertise. 
The underlying physics are nontrivial.
The interface requires microsecond synchronization of all ultrasonic transducers to ensure a correct sound field.
It also requires exact calibration of all physical parts.
Such systems are also difficult to debug. 
In case of problems, the only observable effects are usually the levitating particle dropping or shooting out in an uncontrolled way.
The sources of such problems can be various.
It could be a bug in the software calculating the phase shift for each transducer.
It could be a bug in the underlying libraries, firmware or hardware controlling the ultrasonic transducers.
Finally, it could be physical effects such as reflections of sound waves off neighboring surfaces.
The physical and technical challenges potentially lead to a lack of applications, progress and replication. 

Further, when developing such systems, it is uncertain how far one is from the performance ceiling.
If one would invest in higher framerates, better synchronization, better transducers etc., how much would user performance improve? 
Would there be a noticeable improvement in the user experience when interacting with the upgraded interface?
Lack of this knowledge leads to missed opportunities in some areas and futile iteration in others.


To address these problems, we present the {\it Levitation Simulator}, a prototyping tool aiming to facilitate levitation research and content creation for levitating interfaces.
Instead of having to build an ultrasonic levitation apparatus, designers, artists, software developers and researchers can develop applications and interaction techniques, and perform evaluations and user studies in Virtual Reality (VR).
The volumetric nature of levitation interfaces is preserved in VR, so a comparable visual experience can be provided to the user. 
Only after the development has converged, the resulting system can be validated using a real levitation apparatus, possibly built by another group.

We derive a model of a particle levitating in an ultrasonic field from first principles.
Thus the {\em Levitation Simulator} is able to physically accurately simulate the dynamics of levitating particles.
We validate the Levitation Simulator through a series of two user studies.
First we conduct a Fitts' Law - type pointing study in the {\em Levitation Simulator} and on the real prototype, to obtain a quantitative measure of the pointing performance that can be achieved with both systems. 
Next, we develop two levitation minigames in the {\em Levitation Simulator}, which are then implemented on the real prototype.
In the second user study, User Engagement levels when playing the games on the real and virtual interface, are compared. 
With this study, we want to better understand how users engage in the interaction with levitating matter in the simulator and in the physical world, and what are the most frequently observed differences and similarities.

In summary, the contributions of this paper are:
\begin{itemize}
\item A model that describes the dynamics of movement of a levitating particle in an acoustic field. 
\item An interactive simulation of the levitating interface in Virtual Reality which exhibits the dynamical properties of the real interface - the {\em Levitation Simulator}. 
\item A Fitts' Law pointing study involving aimed movements of a 3D levitating particle between two spherical 3D targets, on the real prototype and in the {\em Levitation Simulator}.
\item A user experience study comparing user engagement levels when playing interactive levitation games, on the real prototype and in the {\em Levitation Simulator}.
\end{itemize}

More generally, we believe that the {\em Levitation Simulator} is a good example to promote modelling and simulation of user interfaces (UIs) in HCI.
Nowadays, the dominant method to develop user interfaces is (physical) prototyping.
Physical prototyping , however, can be difficult, time-consuming, and expensive to build, limiting the number of design iterations.
In these cases, modelling and simulation of the UI can help to increase the number of design iterations at lower costs.

\section{Related Work}
\subsection{Acoustic Levitation}
Ultrasonic levitation interfaces use acoustic radiation force to counteract gravity and trap small objects in mid-air.
This effect can be achieved by using phased arrays of ultrasonic transducers emitting the appropriate phase to create acoustic nodes in mid-air, where the objects are trapped.  
Typically, these traps are generated by creating an acoustic focus.
To create the focus, the phases for each transducer are computed, such that, all acoustic waves constructively interfere at the position, where the levitation trap should be generated.

Even though most research focuses on the technical implementation, several concepts concerning applications for levitating interfaces have been introduced. 
For example, a game where acoustically transparent structures (e.g. tubes made of metallic mesh) are passed around a particle levitating in a standing wave levitator, similar to BigLev~\cite{tinylev}, is presented in~\cite{tangibles}. 
\textit{Floating Charts}~\cite{floating} is a modular display, where levitating particles are used to encode data points on a dynamic chart in mid-air.
An example of a levitating particle being used to trace hiking routes and annotate summits across a model of a mountain range, is shown in~\cite{EnhancingPhysicalObjects}.
The \textit{Pixie Dust}~\cite{pixiedust} system proposed using levitating particles in combination with a projector to form 2D graphics. 
The system can be employed to create raster and vector graphics in mid-air (e.g. logos), as well as to animate physical objects (e.g. products in a store window).
\textit{LeviProps}~\cite{leviprops} are tangible structures composed of an acoustically transparent lightweight fabric and levitating beads as anchors. 
LeviProps can be used as free-form interactive elements and as projection surfaces.
A volumetric acoustophoretic display, where a levitated particle is rapidly displaced, while being illuminated to create 3D shapes, is presented in~\cite{fushimi}.
Hirayama et al.~\cite{hirayama} propose a levitating volumetric display that can simultaneously deliver visual, auditory and tactile content, using acoustophoresis as the single operating principle.




Recently, new interaction techniques allowing users to manipulate levitating particles almost in real time, using gestures, have been developed.
With the \textit{Point-and-Shake}~\cite{pointandshake} method, users are able to select levitating objects by pointing their finger at them.
Visual feedback is provided in the form of a continuous side-to-side (\textit{shake}) movement.
\textit{LeviCursor}~\cite{bachynskyi18} is a method for interactively moving a levitating, physical cursor in 3D with high agility. 
The user controls the levitating cursor with finger gestures. 
Freeman et al.~\cite{EnhancingPhysicalObjects} propose employing these techniques to use the levitating particle as the user representation in interactive applications, such as, to explore landmarks in a miniature model.

\subsection{VR Prototyping}
Prototyping in Virtual Reality is a standard practice in many industries, in particular automotive and aerospace.
A recent survey of the use of Virtual Prototyping in industry is provided by Berg and Vance \cite{berg2017}.
Gomes de Sa and Zachmann \cite{gomesdesa} explore Virtual Prototyping in the Automotive Industry. 
Seth et al. \cite{seth2011} provide a comprehensive review of Virtual Prototyping techniques for assembly processes. 
From the HCI perspective, comparably less studies involving Virtual Prototyping have been conducted.
Aromaa and V\"a\"an\"anen \cite{aromaa} conducted a comparative study of AR vs. VR prototyping for ergonomics evaluation of a maintanance panel of a rock crusher machine.

\subsection{Summary}
In general, the number of applications that have been developed for ultrasonic levitation interfaces has been very limited. 
To our knowledge, no interactive content involving complex actions and information, tailored for levitation displays has been developed.
We are also not aware of formal studies investigating how users interact with more complex content or applications.
With our virtual prototyping tool for levitation, the {\em Levitation Simulator}, we hope to bridge this gap and make prototyping, testing and evaluation of levitation applications accessible to a much wider audience.

\section{Real Prototype}
Our physical ultrasonic levitation system generates the acoustic field that applies forces to a particle, which levitate and move it. 
A motion capture system provides real-time feedback concerning the position of the particle in the levitation volume, the position of the users' fingertip and any optical markers in the observable volume (e.g., the ones defining a rigid body of a controller).

The system uses two rectangular arrays of \SI{14x9}{} ultrasonic transducers each.
The arrays oppose each other and generate a sound field in a levitation volume of \SI{14x9x10.6}{\centi\metre}.
Controller boards and transducer arrays are manufactured by Ultraleap Ltd.
We wrote an application in C, which receives input from the motion capture system.
The application generates the new levitation trap position, as well as runs the experiment. 


\section{Model of the Movement of a Levitating Particle}
In this section, we present a model of the movement of a levitating particle in an ultrasound field. 
We obtain a description of the force acting on the levitating particle by adopting the description by Marzo et al. \cite{marzonature}.
That model is based on a theory from acoustofluidics, which describes the acoustic force acting on small particles in an ultrasound field using the Gor'kov potential \cite{bruus}.

\begin{figure}[t]
\centering 
\includegraphics[width=0.8\columnwidth,]{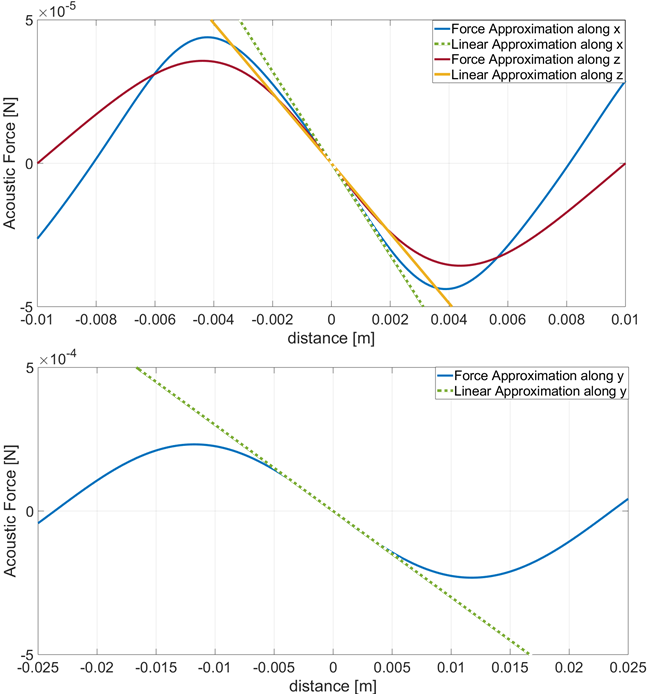} 
\caption{Decomposition of the acoustic force in each spatial dimension. Linear approximation in the vicinity of the levitation trap (dashed green and yellow line).} 
\label{fig:lin_app}
\end{figure}

Figure \ref{fig:lin_app} shows the individual force components, in each spatial dimension.
We observe strong acoustic force vectors in the y-direction, pushing the particle from above and below towards the center of the trap.
Also in the vicinity of the levitation trap, in all three directions, the force profile is of a sinusoidal shape, meaning that the acoustic force acts as a restoring force, pushing the particle towards the center of the trap, while air friction dampens its motion. 
The force profiles indicate the maximal distance we can push a particle away from the trap center and it would still be pulled back at its original (equilibrium) position, i.e. the trap size (ellipsoid with diameters $d_{x}\approx$ \SI{16}{\milli\metre}, $d_{y}\approx$ \SI{5}{\milli\metre} and $d_{z}\approx$ \SI{20}{\milli\metre}). 
The maximal force in the x-direction is \SI{4.4e-5}{\newton}, \SI{2.2e-4}{\newton} in the y and \SI{3.5e-5}{\newton} in the z-direction.
The maximal force that the acoustic field can exert to the levitating particle sets the physical limit for its velocity and acceleration.

Note that the acoustic force in the y-direction is an order of magnitude higher than the two horizontal forces, implying that the levitating particle can achieve higher accelerations and velocities in this direction compared to the other two directions. 
This result can have implications for performing pointing tasks in 3D on the levitating display.
While in pointing we often assume homogeneity of movement in all directions, this assumption is clearly violated for ultrasonic levitation. 

Let us denote the displacement of the levitating particle from the trap by $x$. The balance of forces using Newton's second law gives 
\begin{equation} \label{eqn_force_balance}
F=ma=F_{ext}+F_{a}-F_{drag},
\end{equation}
where $F$ is the net force acting on the particle, $F_{ext}$ is the external force, e.g., from gravity or wind; $F_{a}$ is the acoustic force, produced by the ultrasonic transducers; $F_{drag}$ is the air resistance, $m$ denotes the mass of the levitating particle and $a$ its acceleration. 
We assume that when the particle moves through air, there is no turbulence (i.e. low Reynolds number), so the drag force can be modeled to be proportional to the velocity.
We set $F_{drag}=c\frac{dx}{dt}$, where $c$ is the drag constant, that describes the decay of oscillations after the particle has been disturbed. 
The gravitational force acting on the levitating particle is two to three orders of magnitude, depending on the particle size, smaller than the acoustic force, so it can be neglected in the model. 
In addition, we make the assumption that no other external force acts on the system, i.e. $F_{ext}=0$. Hence, we obtain $F=ma=F_{a}-c\frac{dx}{dt}$, which by setting $v=\frac{dx}{dt}$, can be rewritten into the system of equations
\begin{align} 
	\dot{x} &=  v \\ 
	\dot{v} &= \frac{F_a-cv}{m}.
\end{align}

The theory of the acoustic radiation force, stemming from the scattering of acoustic waves on a small particle, has been extensively discussed in~\cite{bruus}. 
Without going in detail, here we only present the fundamental acoustic force expression
\begin{equation} \label{eqn_force_gor}
F_{a}=-\nabla U=-U_{x}-U_{y}-U_{z},
\end{equation}
where the force is described as a gradient of the Gor'kov potential $U$~\cite{gorkov}, which in turn is dependent on the complex acoustic pressure and its spatial derivatives. 

We note that the analytical expression of the acoustic force  (Eq.~\ref{eqn_force_gor}) is too complex for our modeling purposes, so we  approximate it, in the neighborhood of the trap, by a linear function i.e. $F_a=bx$ (see Figure \ref{fig:lin_app}).
We use Taylor series expansion, to identify the vector of coefficients $b=[0.016, 0.26, 0.011]$.

The diameter of the levitating particle is measured to be $d=$ \SI{0.002}{\meter} and the material density of expanded polystyrene is estimated to be $\rho_{EPS}=$ \SI{25}{\kilogram\per\meter\cubed}, thus we obtain a particle mass of $m\approx$ \SI{1.05e-7}{\kilogram}. 
The linear drag constant $c\approx$ \SI{9.42}{\kilogram\per\second} was empirically obtained.

\begin{figure}[h]
\centering 
\includegraphics[width=0.85\columnwidth,]{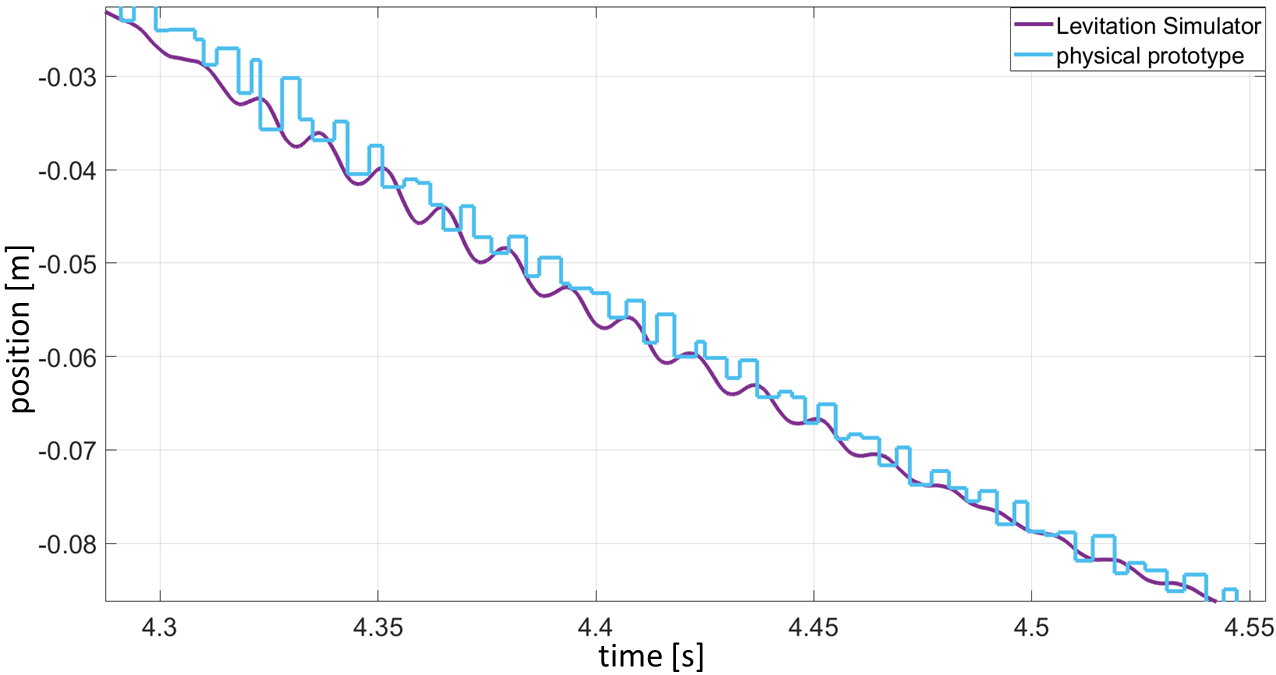} 
\caption{Comparison of output data of the {\em Levitation Simulator} to the measured data from the real prototype.} 
\label{fig:compare}
\end{figure}

According to our experiences, the particle movement in VR appears to be indistinguishable from the particle movement in the real prototype.
A comparison of simulated particle movements to measured data from the real prototype is provided in Figure~\ref{fig:compare}. 

\section{The Levitation Simulator}
The {\em Levitation Simulator} consists of an interaction and a simulation module.
The interaction module is implemented in C\# within the Unity Game engine.
Users can interact with the simulated levitation interface either via a motion capture system or via a VR controller.
The levitated particles move realistically, as they would with a physical levitation system.

Applications and experiments can be implemented through Unity scripts. 
One script we provide, records the details of the movement, including particle and finger position per frame with timestamps, to a CSV file.
Another script reads the input, e.g., from the OptiTrack Streaming Client via NatNet or from the VR controller.
It transforms the input, e.g., by applying C:D gain, and sends the new trap position via UDP to the simulation module.
It receives the new particle position via UDP from the simulation module and renders the scene.
The entire system runs at 90 FPS, synchronized with the update rate of the HMD.

\begin{figure}[h]
\centering 
\includegraphics[width=\columnwidth,height=2.4cm]{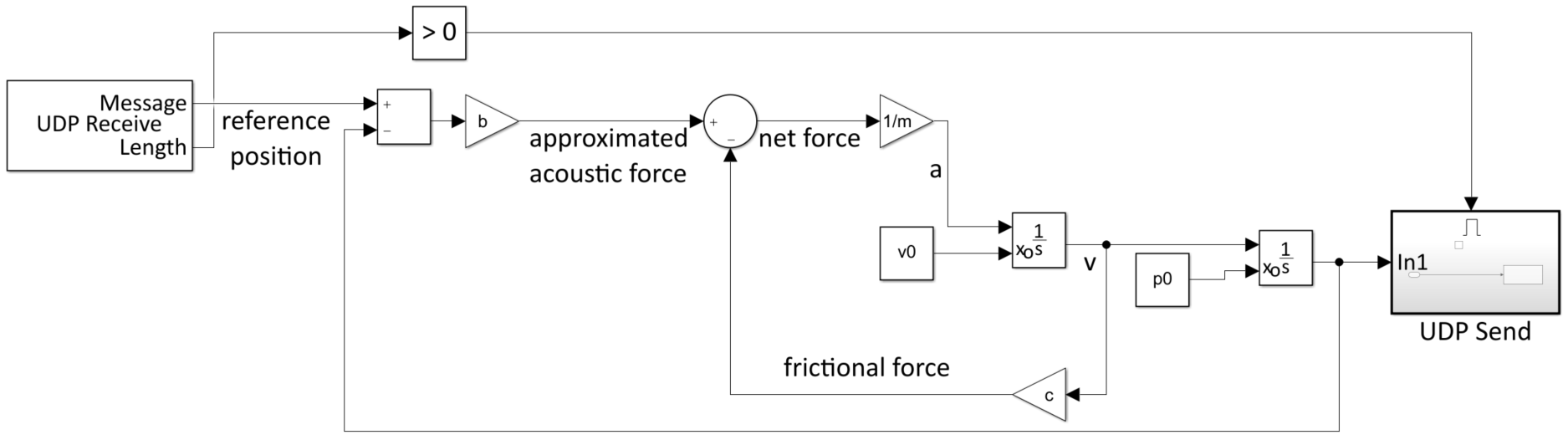} 
\caption{Model of the dynamics of the levitating particle in Simulink.} 
\label{fig:sim_mod}
\end{figure}

The simulation module (Figure \ref{fig:sim_mod}) is implemented in the {\em Matlab Simulink} environment.
It reads the new trap position from a UDP socket.
It maintains the velocity and position of levitated particles in two integrators.
From the new position of the trap and the current position of the particle, it calculates the force that is acting on the particle.
To calculate the force, it uses the model described above.
It then simulates the effect of that force on the particle.
The new particle position is transmitted to the interaction module via UDP.
In order to minimize latency, we use no UDP buffers.
We use a variable-step solver (Runge-Kutta (4,5)).

We currently use an HTC Vive Pro HMD and Optitrack Motion Capture System with Prime 13 cameras. 

\begin{figure}[]
\centering 
\includegraphics[width=0.9\columnwidth,]{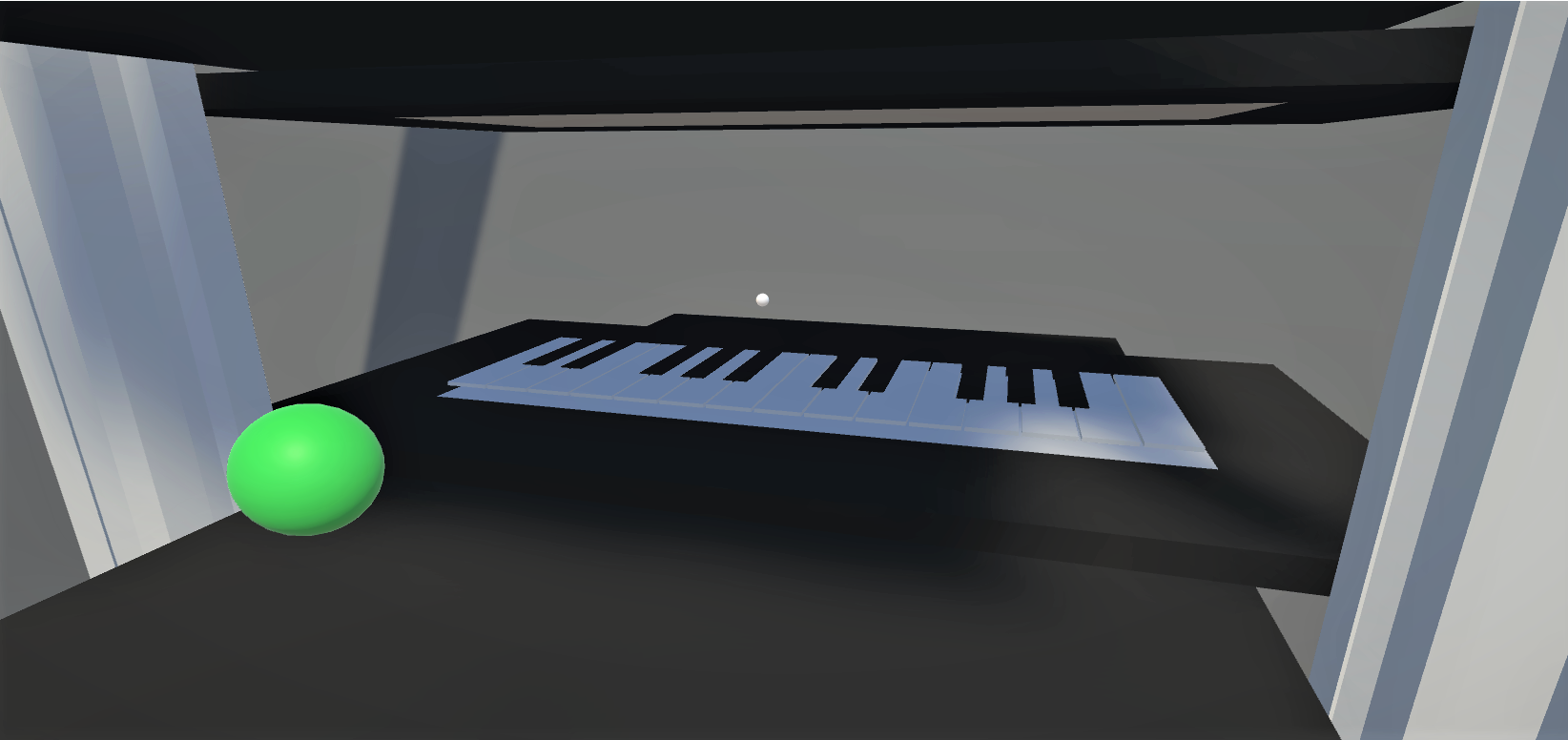} 
\caption{The ``Levitating Piano'' is an application developed in the {\em Levitation Simulator}. The user has a $3$~mm wide retroreflective marker placed on their fingernail. With their finger (green sphere), they control the levitating particle (white sphere) in the levitating display. In this way, they can play a virtual levitating piano, made of acoustically transparent fabric. Whenever the levitating particle touches the fabric, the corresponding piano tone sounds (see accompanying video).} 
\label{fig:piano}
\end{figure}

This design makes the {\em Levitation Simulator} flexible and easy-to-operate, and makes changes to the underlying model trivial to implement.
In particular, we can swap models of particle behavior to simulate different levitation apparatuses, different levitated particles, or different techniques to generate the sound field. 
In addition, it allows us to simulate and test application ideas that would take months to implement on the real prototype, such as the Levitating Piano (see Figure \ref{fig:piano}).

The {\em Levitation Simulator} is simple to use, if one does not want to change interface or simulation. 
Extending it might require some expertise in Unity and/or Simulink, depending whether one wants to change the interface or the simulation.
As such skills are available in the HCI community, we believe that the {\em Levitation Simulator} will be easy to work with and extend for research groups in the HCI community. 
For users that do not wish to experiment with different models of the movement dynamics, we provide a stand-alone Unity version of the {\em Levitation Simulator}, where the model in Figure~\ref{fig:sim_mod} is integrated into the Unity environment.

\section{Pointing Study}
We are interested in how far the interaction with the real and virtual prototype enables similar pointing performance, and if not, what the differences are.
Further, we want to investigate whether the control of a levitating cursor with the real and virtual prototype, is perceived similarly by users.
Thus we evaluate the performance achieved by interacting with the \textit{Levitation Simulator}, compared to the real prototype, in a Fitts' Law type repetitive pointing study. 

\subsection{Experimental Design}
Participants performed repetitive aimed mid-air movements between two three-dimensional, spherical targets of varying size.
We used a within subjects experimental design with two interface conditions (real and VR), two movement directionality conditions (left-right and front-back), and three index of difficulty conditions (target amplitude \SI{5}{\centi\meter} and three target diameters: \SI{4}, \SI{8} and \SI{16}{\milli\meter}). 
The indices of difficulty covered in the experiment are 2.04, 2.85 and 3.75.
The range of possible indices was constrained by the properties of the real prototype: the distance between the targets was constrained by the sound field region allowing stable levitation and agile control, the target size was limited by the size of the cursor and the persistent oscillations of the particle within the trap.
The participants performed a set of equivalent Fitts' law reciprocal pointing tasks, both on the real prototype and in the {\em Levitation Simulator}.
The order of interfaces was counterbalanced by randomization, and the order of conditions on each interface was fully counterbalanced by a Latin square.
The participants performed $70$ aimed movements for each condition.
We recorded both, the position of the fingertip marker and the levitating particle, on a frame-by-frame basis, as well as the timings of the auditory feedback and the individual aimed movement durations.

\subsection{Participants}
Twelve healthy participants (4 females), aged between 20 and 39 (mean 27.25, SD 5.15), were recruited for the experiment. 
They all had normal or corrected-to-normal vision. 
All preferred to control the levitating particle with their right hand. 
Before the experiment, they were provided with basic information about the study and signed a consent form.
The experiment was approved by the Ethical Committee of the University of Bayreuth and followed the ethical standards of the Helsinki Declaration.
All participants received monetary reimbursement for their participation.

\subsection{Apparatus}
\begin{figure}[tbh]
\centering 
\includegraphics[width=\columnwidth,]{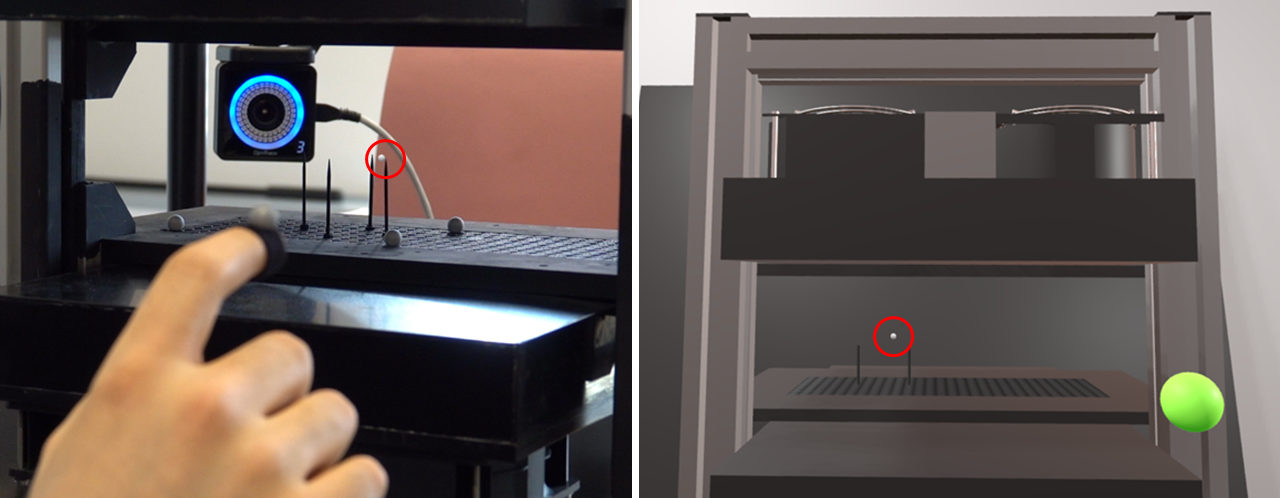} 
\caption{Real acoustic levitation interface (left) and {\em Levitation Simulator} (right). For orientation, the green sphere indicates the position of the user's fingertip.}  
\label{fig:setup}
\end{figure}

In the experiment, we have used the real prototype and the {\em Levitation Simulator}, described in the previous sections, as the basis for control of the levitating particle.
The experimental setup is illustrated in Figure~\ref{fig:setup}.
The targets for the Fitts' law tasks were implemented as thin pointy black sticks, placed between the transducer arrays, to reduce interference with the acoustic field and the optical tracking system.
In the {\em Levitation Simulator}, equivalent sticks were placed in the scene.
Internally, in both systems, the targets were represented as spheres of three different sizes, centered at the tips of the sticks.
While the actual spheres were invisible, the system produced an auditory feedback as soon as the levitating particle entered the target sphere. 
Both applications processed and recorded the motion capture data describing the movement of the finger and the position of the (real or simulated) levitated particle. 
Participants could control the particle motion in 3D with their fingertip, using a control-to-display ratio of 3. 

\subsection{Procedure}
Before beginning the experiment, the participants read basic information about the study and a description of the tasks.
The experimenter provided clarifications, if the participants had any questions.
The participants sat on a chair in front of the apparatus.
In the VR condition, they were asked to wear the HTC Vive Pro and adjust it according to their comfort.
They were asked to take a comfortable position, adjust the chair height and location to ensure good visibility of the targets both in the {\em Levitation Simulator} and the real prototype.
A retroreflective marker of \SI{9}{\milli\meter} diameter was attached to the fingertip of the index finger of the dominant hand of each participant.
In the real condition, the experimenter placed a white polystyrene particle into the levitation volume.
We started with an exploration phase, in which the participants explored the interaction with the interface and moved the levitating particle in different directions for approximately \SI{30}{\second}.
We asked the participants to place the particle at each of the tips of the target indicators, as accurately as possible.
We then calibrated the target locations according to the participants' individual perception of the target.
Next, we conducted a training phase for each condition. 
The participants were instructed to move between the targets as quickly and accurately as possible, for $20$ repetitions.
In the experiment, participants were asked to perform $70$ repetitions for each condition. 
During the experiment, our software was continuously recording the time-stamped 3D position of the levitated particle at each frame.
We also recorded the time stamps when the user reached each target and was notified by the sound. 
After each condition, the participants were asked to take a short break. 
After the completion of all tasks in the first interface condition, the participant was moved to another room to continue with the other condition.
Subsequent to completing all pointing tasks in each environment, the participants filled in the NASA Task Load Index questionnaire~\cite{hart_tlx}.
At the end, we performed a semi-structured interview with each participant.

\subsection{Analysis}
We performed two types of analyses on the collected data.
We used Fitts' law modeling for comparing the performance achieved on the real prototype and in the {\em Levitation Simulator}.
We used statistical analysis for the impact of conditions on movement time, as well as for the NASA TLX data. 

As preprocessing, we average the movement time values per condition.
The movement times were not normally distributed according to the Kolmogorov-Smirnov test ($p > 0.05$), so we opted for a non-parametric statistical analysis. 
We performed a Friedman test to compare movement times between the interfaces and IDs. 
We use a Wilcoxon signed-rank test to analyze the NASA TLX data.

The Fitts' law modeling was performed according to the common HCI practice~\cite{SOUKOREFF2004751}.
While the pointing tasks and the targets were represented in 3D space and multivariate Fitts' law models could be applied~\cite{Grossman:2004:PTT:985692.985749}, with spherical targets they are equivalent to the application of Fitts' law in the Shannon formulation:
\begin{equation}
ID=\log_2\left(\frac{D}{W}+1\right)
\end{equation}
\begin{equation}
MT=a+b\times ID
\end{equation}
\begin{equation}
TP=\frac{1}{y}\sum_{i=1}^{y}\left(\frac{1}{x}\sum_{j=1}^{x}\frac{ID}{MT}\right)
\end{equation}
where $ID$ is index of difficulty, $D$ and $W$ are the movement amplitude and target width respectively, $MT$ is movement time, $a$ and $b$ are regression coefficients, and $TP$ is information throughput.
In the preprocessing step, we discard the first 20 movements from each trial, as the participants were adjusting to the new size of the invisible target.
We then compute $ID$ and average $MT$ for each trial. 
Before fitting the regression lines, we group all trials according to the $ID$ and compute average $MT$ and $ID$ for each group.
Similarly, we use the group values to compute the movement performance $TP$.

\subsection{Results}
\subsubsection{Pointing performance}
The results of the Friedman test show no significant differences in movement times between the conditions.
We can however observe some trends in the data, which we describe in the following.
We can see a trend, according to which with increase in index of difficulty, the movement time increases faster in the VR condition than in the real condition.
For increasing IDs of $2.04$, $2.85$, and $3.75$~bits, the average movement time in the real condition is $0.665$, $0.823$, and $1.028$~s, respectively, compared to the respective movement times of $0.679$, $0.876$, and $1.253$~s, in the VR condition. 

When considering the effect of movement direction, we observe another trend.
With increasing index of difficulty, the movement time increases faster in the left-right condition than in the front-back condition, in the {\em Levitation Simulator}.
For the same increasing sequence of IDs as above, the average movement times in the front-back condition are $0.686$, $0.863$, and $1.206$~s.
In the left-right condition, the movement times are $0.671$, $0.890$, and $1.301$~s.

We can see no such trend in the real condition.
Both left-right and front-back movements exhibit almost the same movement times for each index of difficulty. 
There is only a small constant difference.
For the same increasing sequence of IDs as above, the average movement times in the front-back condition are $0.674$, $0.814$, and $1.038$~s.
In the left-right condition the movement times are $0.656$, $0.830$, and $1.019$~s.


\begin{figure}[h]
	\centering
	\includegraphics[width=0.9\columnwidth,]{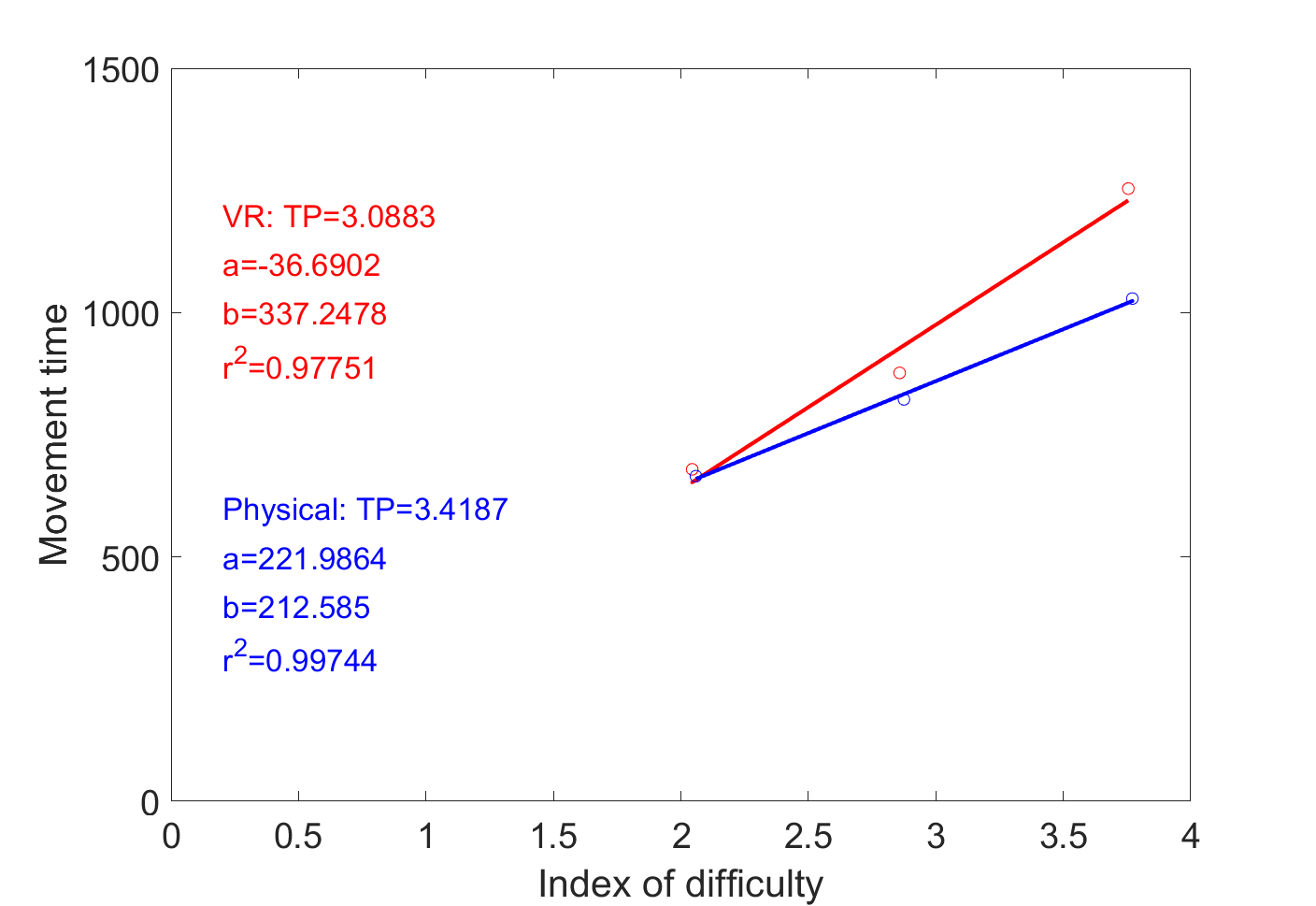} 
	\caption{Fitts' law models for {\em Levitation Simulator} vs. physical apparatus conditions. The circles represent the average MT corresponding to the three IDs, in each condition.}
	\label{fig:fittsVrLv}
\end{figure}

We computed Fitts' law models for the different interfaces, and further considered the effects of movement direction for the whole data.
The Fitts' law models comparing the real and VR condition are shown in Figure~\ref{fig:fittsVrLv}.
Models in both the {\em Levitation Simulator} and real prototype conditions, as well as for all four condition combinations provided a good fit for the data with $R^2>0.97$.
The models reflect the trends in the movement time between conditions described in the previous paragraph.
We can see that both conditions are characterized by similar movement times and similar throughput, with the throughput of the real prototype being slightly higher.
In the VR condition, the throughput is $3.08$~bits/s.
In the real condition, the throughput is $3.41$~bits/s.
The regression line of the {\em Levitation Simulator} condition is steeper ($b=337$) than for the real condition ($b=212$).
Movement time increases faster in the VR than in the real condition.

\begin{figure}[tb]
	\centering
	\includegraphics[width=0.9\columnwidth,]{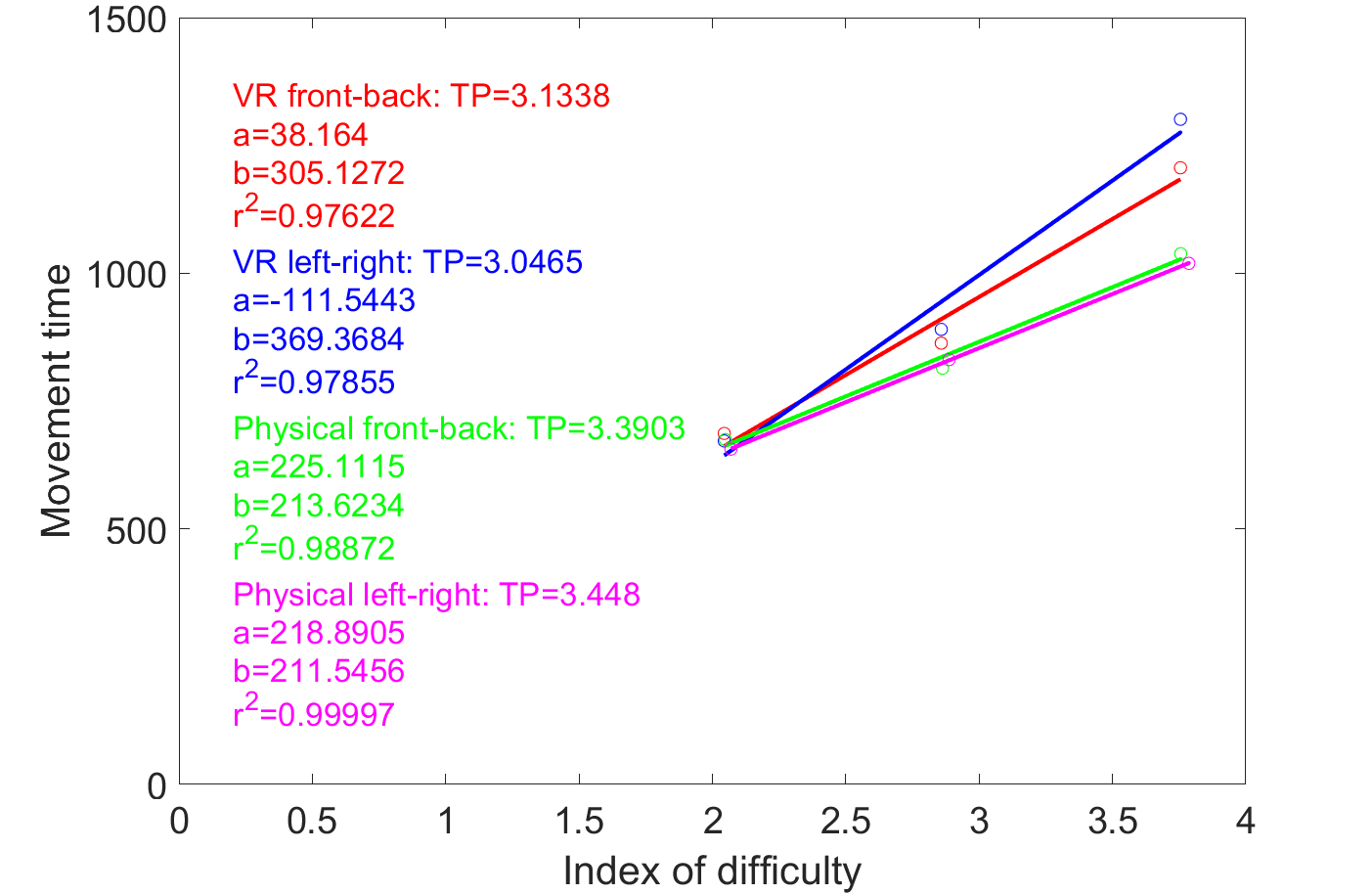} 
	\caption{Fitts' law models for each combination of interface and direction conditions. The circles represent the average MT corresponding to the three IDs, in each condition.}
	\label{fig:fittsConditions}
\end{figure}

Figure~\ref{fig:fittsConditions} shows the Fitts' law analysis, split by interface and direction.
We can observe that movement direction does not affect the movement times in the real prototype condition (slope $b=211$ in left-right condition and $b=213$ in front-back condition).
It does however affect the movement times in the {\em Levitation Simulator} condition.
The movement times of the left-right movements in the VR condition grow faster with increasing index of difficulty, than the movement time  of the front-back movements ($b=369$ for left-right vs. $b=305$ for front-back conditions).
%
%
%

\subsubsection{Task Load Index (TLX)}
\begin{figure}[h]
\centering 
\includegraphics[width=0.9\columnwidth,]{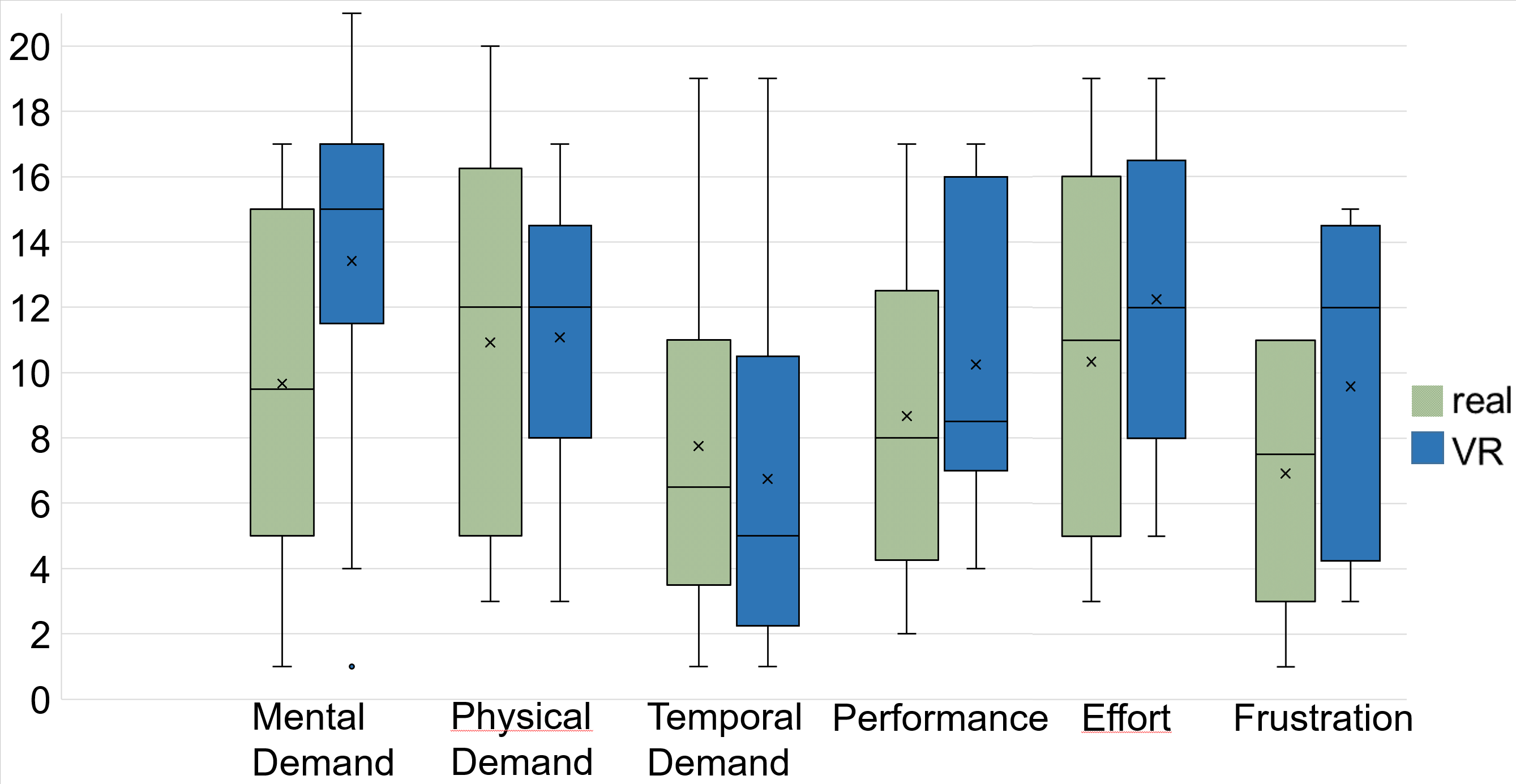} 
\caption{Boxplot of the scores on the NASA TLX factors for the Fitt's Law pointing task, in the real and VR condition.} 
\label{fig:TLX}
\end{figure}

The results of the NASA TLX questionnaire are shown in Figure~\ref{fig:TLX}. 
Overall, the workload in VR was reported as slightly higher (mean=$10.5$), compared to the workload in the physical environment (mean=9.0).
The Wilcoxon signed-rank did not show any significant difference in the mental demand ($p=0.066$), physical demand($p=0.92$), temporal demand ($p=0.089$), performance ($p=0.12$), effort ($p=0.37$) and frustration ($p=0.15$) scores, for the real and VR conditions. 

\subsubsection{Observations and Interviews}
During the experiment, we observed different strategies to complete the pointing task.
In tasks with a low index of difficulty, the participants relied more on the visual cues when trying to hit the target.
With a high index of difficulty, the exact depth of the target was difficult to see, so some participants relied more on their muscle memory.
They tried to memorize and reproduce the gesture that produced a successful target hit.
Some participants reported to have used the collision of the levitating particle with the target indicators, in both the VR and the real condition, as a visual cue.
On the physical apparatus, some participants also used the increase in oscillations of the particle around the physical targets as additional visual feedback.
These oscillations occur as the target indicators slightly disturb the sound field.
We did not model this effect in the {\em Levitation Simulator} yet. 

Almost all participants reported in the interviews, that the interaction felt very similar in both conditions.
In both conditions, they felt high control over the levitating particle.
There was an interesting difference between participants who had experience in VR and those who had not.
Those who had previous VR experience, were very impressed by the physically realistic movement of the particle in the {\em Levitation Simulator}.
They had not had such a realistic experience in VR before.
In contrast, those who had not experienced VR before, were rather underwhelmed by the overall realism of the {\em Levitation Simulator} experience.
Apparently, they had much higher expectations of the realism of VR experiences in general. 

On the real levitation interface, all participants showed amazement over real levitation.
They expressed this with phrases such as that when they controlled the particle with their finger they felt as a \textit{wizard with a magic wand}.
This might be due to the fact that none of our participants have experienced a levitation interface before.
This amazement was not expressed with the {\em Levitation Simulator}. 

The most frequent criticism of the {\em Levitation Simulator} was the limited quality of the depth cues in VR.
When performing the experimental task, participants clearly had difficulty judging the depth of the levitating particle relative to the targets.
This was the case particularly for the left-right movements.
Participants found this particularly frustrating when they compared it with the real prototype, which provided all natural depth cues perfectly.
Some participants also complained about the quality of the resolution of the HMD Vive Pro headset.

In both conditions, the participants reported that they felt they could perform the task better if the targets were visible. 
Participants made further suggestions to increase the levitation volume, to add visible targets and to improve the depth perception in VR. 

\section{User Engagement Study}

The objective of our second user study was to investigate differences and similarities in the way users engage and interact with the same applications, presented in the {\em Levitation Simulator} and on a physical levitation apparatus. 
For this purpose, two different games for levitation interfaces, a ball-and-racket game and a first person shooter, were developed in the {\em Levitation Simulator}, and then implemented on the real prototype.
The ideas for the games were generated in a brainstorming session. 
The exact game design and movement parameters were determined in the {\em Levitation Simulator} in an iterative testing process with pilot participants. The biggest challenge was to find the optimal bead velocity, such that the game is challenging enough for the player, but not too difficult so they become frustrated and quit. 
With the {\em Levitation Simulator} it was possible to efficiently test many velocities and collect user responses. 

\begin{figure}[tbh]
\centering 
\includegraphics[width=\columnwidth,]{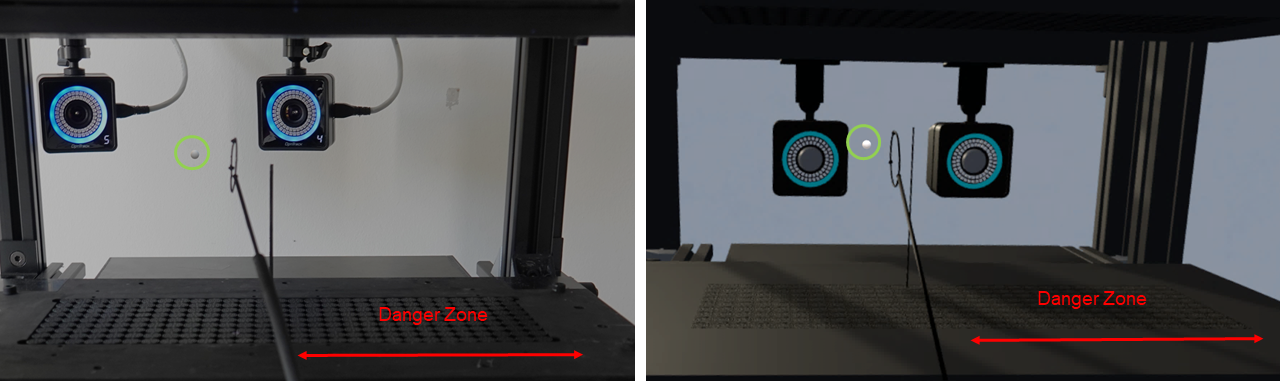} 
\caption{In the BeadBounce game, the player has to prevent the levitating particle (marked with a green circle) from bouncing off into the \textit{danger zone} (red arrow), by hitting it with the racket.}  
\label{fig:BeadBounce}
\end{figure}

The first game developed was {\it BeadBounce}.
The goal of the game is to prevent the levitating particle from going into the 'danger 
zone'. 
The game is played in the left half of the levitation interface. 
The right half is the danger zone. 
The entrance to the danger zone is marked by a black pole, positioned at the back of the levitation volume, to allow for free movement of the racket.
During the game, the levitating particle moves in a straight 3D line and bounces off the walls of the levitating volume. 
When the particle starts moving towards the danger zone, the player has to hit it with the racket controller, so that it bounces back to the left (see Figure~\ref{fig:BeadBounce}). 
Initially the particle starts moving with $0.09$~m/s.
When the particle is hit, part of the racket momentum transfers to the levitating particle, hence the particle moves with varying velocity, during the course of the game.
Audio feedback is given whenever the levitating particle bounces off the wall or the racket.

\begin{figure}[tbh]
\centering 
\includegraphics[width=\columnwidth,]{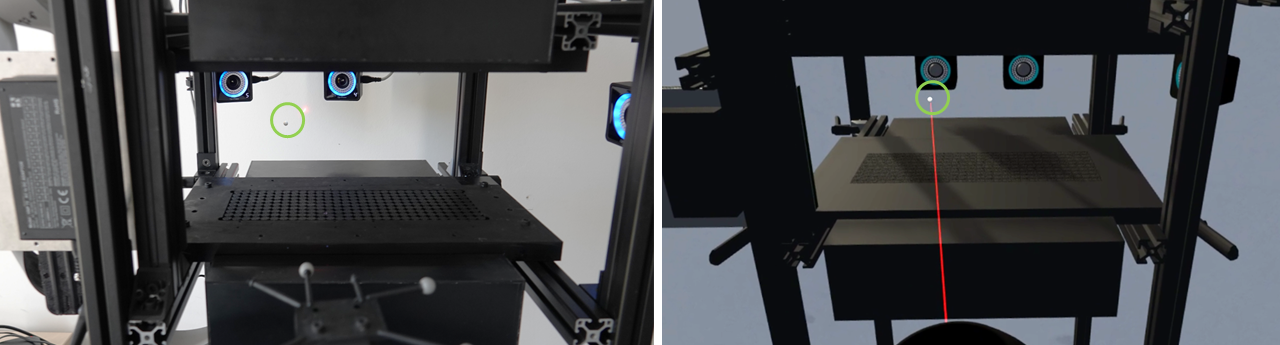} 
\caption{In the LeviShooter game, the player aims at the levitating particle (marked with a green circle), using a lasergun. The goal is to successfully hit the particle, as often as possible, while with every hit, the particle velocity increases.}  
\label{fig:LeviShooter}
\end{figure}

The second game was {\it LeviShooter}.
In this game, the player shoots at a levitating particle with a laser gun controller (see Figure~\ref{fig:LeviShooter}).
The levitating particle moves in a straight 3D line, with an initial velocity of $0.05$~m/s. 
The player needs to aim at it accurately. 
Visual feedback that the gun controller is aiming correctly, is given by the reflection of the laser beam from the particle. 
When the particle is successfully hit, the player has to wait $2$~s until they can shoot again. 
With every hit, the difficulty of the game increases, by increasing the velocity of the particle, by an increment of $0.001$~m/s. 
After ten misses, the levitating particle lowers its velocity again. 
Audio feedback is given whenever the lasergun is fired, and when the particle is successfully hit.

\subsection{Experimental Design}
The user study consisted of a within-groups experimental design, with two interface conditions (real and VR), for each of the two games.
The order of conditions was fully counterbalanced by a Latin square.

To measure user engagement levels, we adopted the long version of the User Engagement Scale (UES)~\cite{obrien}. 
The scale uses four dimension identifiers: Focused Attention - measures the degree of being absorbed in the experience and losing track of time, Perceived Usability - evaluates interface usability, level of control and how demanding the experience was, Aesthetic Appeal - appeal of the interface to the visual senses, and Reward - assesses whether the experience was fun, rewarding and worthwhile.
The UES questionnaire consists of $30$ items, where each is answered on a five-point rating scale.
We randomized the order of items prior to administering.

We also measured how long participants played the games, and conducted semistructured interviews with them. 

\subsection{Participants}
Twenty four participants ($10$ females and $14$ males) aged between $19$ and $32$ (mean $24.17$, SD $3.95$), were recruited for the experiment. 
All had normal or corrected-to-normal vision and no previous experience with levitating interfaces.
Before the experiment started, the participants read basic information about the study and signed a consent form.
The study was approved by the Ethical Committee of the University of Bayreuth. 
All participants received a monetary reimbursement for their participation.

\subsection{Apparatus}
The study was conducted using the {\em Levitation Simulator} and the real prototype, as previously described. 
For the BeadBounce game on the physical apparatus, we designed and 3D printed a racket that can interact directly with the levitating particle within the levitating volume, without disturbing the acoustic field and that can be tracked by the motion capture system.
The racket head is hollow, with a diameter of $3$~cm and frame thickness of $1$~mm.
The racket handle is $40$~cm long with thickness of $6$~mm.
We attached five $3$~mm wide retroreflective markers to the racket handle, to enable optical motion tracking.
In the LeviShooter game, the input device consisted of a 3D printed laser-gun-shaped casing.
We placed a laser pointer and a trigger inside of the casing, and mounted five $9$~mm wide retroreflective markers on top.

\subsection{Procedure}
Upon arrival, participants were given basic information about the study and the rules of the levitation games were explained to them, by the experimenter. 
They were asked to take a comfortable position, either standing up or sitting down, that allowed them a good overview over the levitation volume. 
In the VR condition, participants were asked to adjust the HTC Vive Pro headset to their comfort. 
Then they were given either a racket or a lasergun controller and started playing BeadBounce, or respectively LeviShooter.
The participants were instructed to play the games as long as they liked. 
We recorded the gameplay duration and the score.
If the gameplay exceeded $5$ minutes, we asked them to move to the next condition.
After each condition, the participants took a short break and filled in the UES questionnaire. 
At the end of the experiment, we conducted a semi-structured interview, where we asked participants about their game experience in the {\em Levitation Simulator} and on the real prototype.

\subsection{Analysis}
The UES questionnaire data and the gameplay times were statistically analyzed.
The normality assumption of the UES data was confirmed by performing a Shapiro-Wilk test (p>0.05).
We analyze the data using a paired t-test.
Using a Shapiro-Wilk test, we also checked the normality of the distribution of gameplay times data. 
According to the test, we cannot assume normality for the data obtained in the VR condition of the BeadBounce game ($W=0.75, p<0.01$) and in the VR condition of LeviShooter ($W=0.78, p<0.01$). 
Thus we opt for non-parametric statistical analysis and perform a Wilcoxon signed-rank test.

\subsection{Results}
\subsubsection{User Engagement}

\begin{figure}[b]
	\centering
	\includegraphics[width=0.9\columnwidth,]{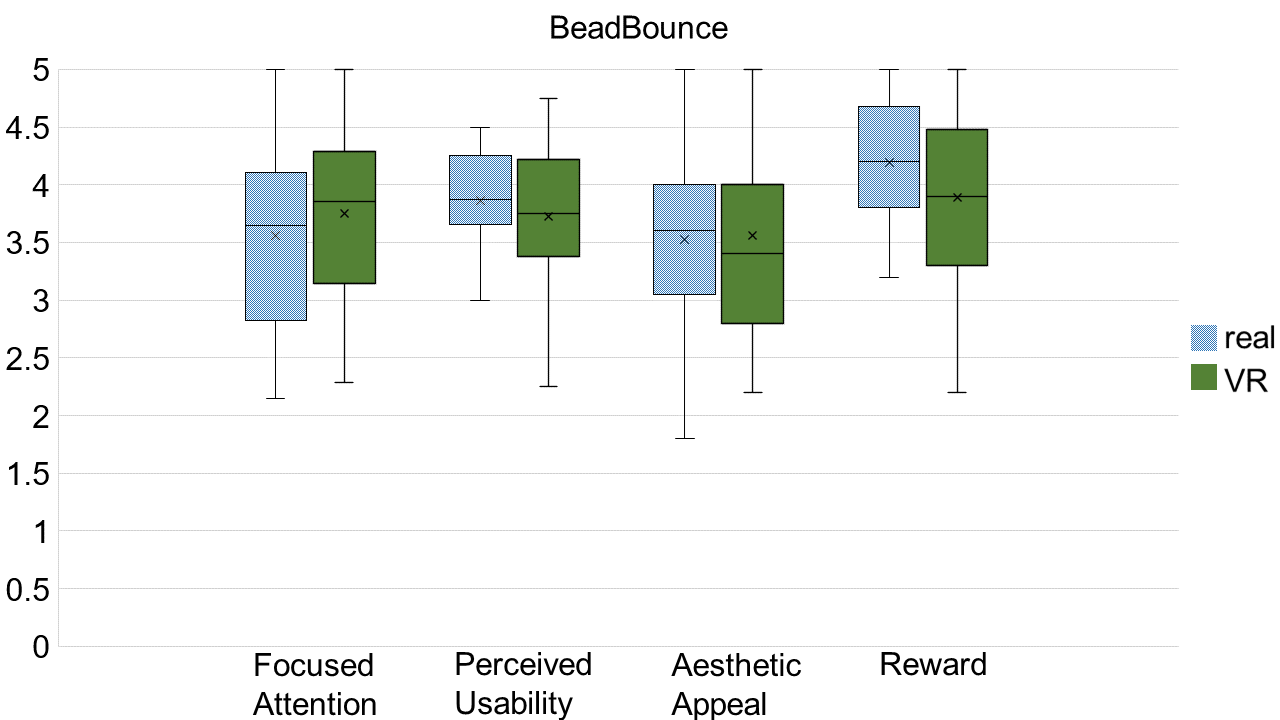} 
	\caption{Scores for the BeadBounce game on the physical apparatus and in the {\em Levitation Simulator}, in each subscale of the UES Questionnaire.}
	\label{fig:UES_BeadBounce}
\end{figure}

The UES scores obtained for the BeadBounce game, in each subscale, are shown in Figure~\ref{fig:UES_BeadBounce}. 
There was a significant difference in the Reward scores for the real (M=$4.20$, SD=$0.56$) and VR (M=$3.89$, SD=$0.70$) conditions; ($t(23)=4.05, p<0.001$).
No significant difference was found in the Focused Attention ($p=0.17$), Perceived Usability ($p=0.32$) and Aesthetic Appeal ($p=0.80$) scores, for the real and VR conditions. 


\begin{figure}[tb]
	\centering
	\includegraphics[width=0.9\columnwidth,]{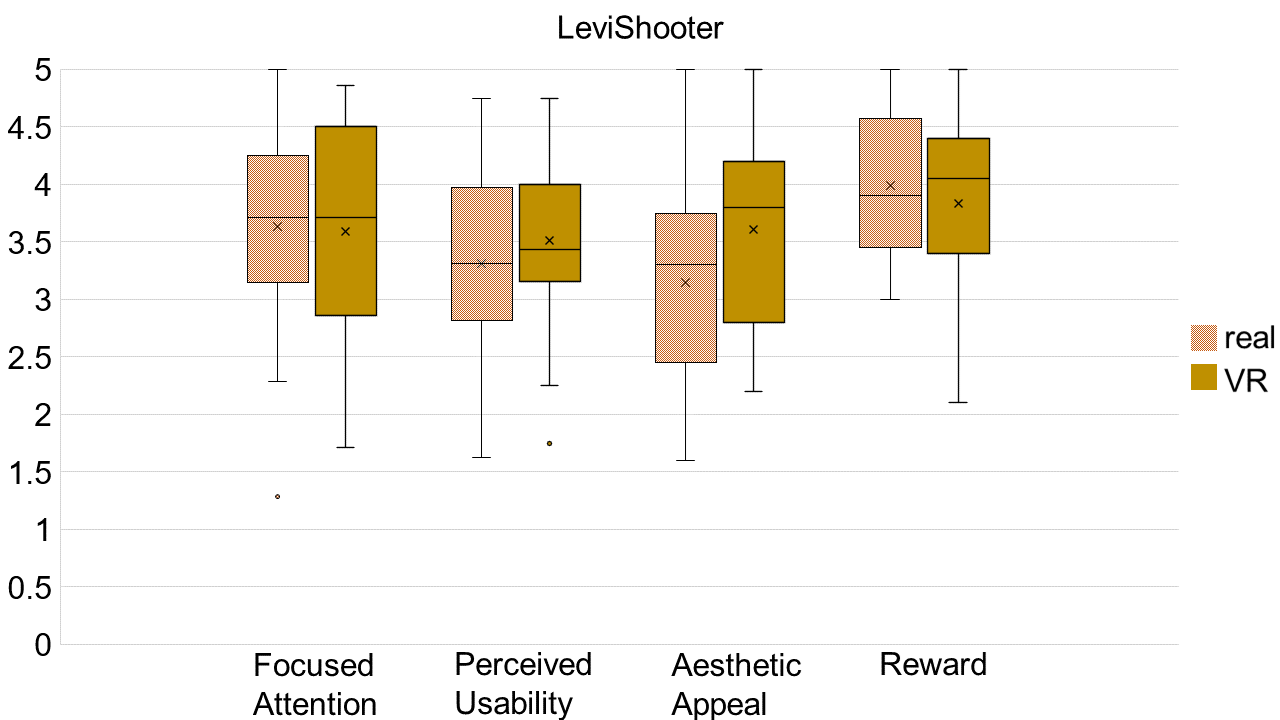} 
	\caption{Scores for the LeviShooter game on the physical apparatus and in the {\em Levitation Simulator}, in each subscale of the UES Questionnaire.}
	\label{fig:UES_LeviShooter}
\end{figure}

The scores in each UES subscale, for the LeviShooter game, are presented in Figure~\ref{fig:UES_LeviShooter}.
There was a significant difference in the Aesthetic Appeal scores for the real (M=$3.15$, SD=$0.84$) and VR (M=$3.61$, SD=$0.82$) conditions; ($t(23)=-3.0731, p=0.005$).
No significant difference was found in the Focused Attention ($p=0.85$), Perceived Usability ($p=0.29$) and Reward ($p=0.40$) scores, for the real and VR conditions. 


\subsubsection{Gameplay Times}
\begin{figure}[tbh]
	\centering
	\includegraphics[width=0.85\columnwidth,]{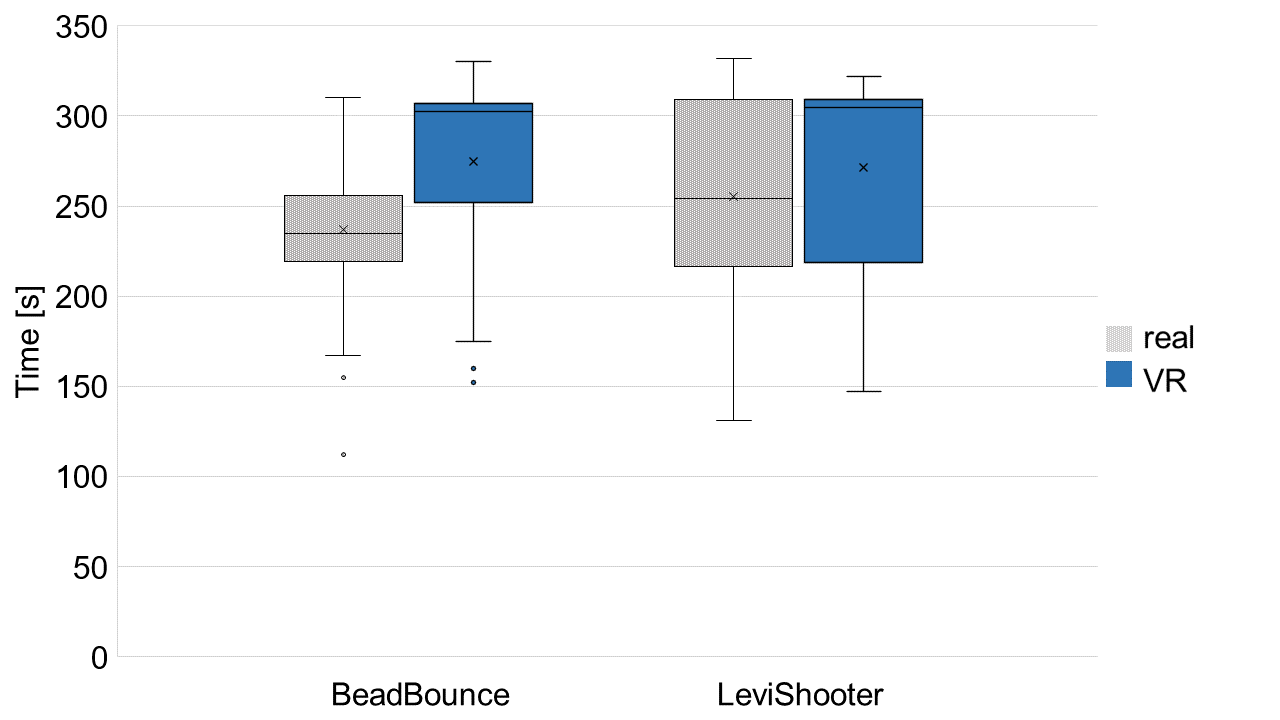} 
	\caption{Boxplot for the gameplay times per game and interface condition.}
	\label{fig:playtime}
\end{figure}

Means and standard deviations for the gameplay times in each condition, are shown in Table~\ref{mean_playtime}. 
A Wilcoxon signed-rank test indicates that participants spent significantly more time playing the BeadBounce game in the VR (Mdn$=302.5$), compared to the real condition (Mdn$=235$); ($p<0.001$).
The scores are presented in Figure~\ref{fig:playtime}.
No significant difference was found for the gameplay times in the LeviShooter game ($p=0.39$). 

 \begin{table}[tbh]
 \resizebox{\columnwidth}{!}{
 \begin{tabular}{|l|c|c|c|c|}
  \cline{1-5}
Game &  \multicolumn{2}{c|}{BeadBounce}& \multicolumn{2}{|c|}{LeviShooter}  \\
 \cline{1-5}
Interface & & & & \\
Condition & real & VR &  real & VR  \\
 \cline{1-5}
Mean & & & & \\
Playtime & $236.92$($\pm47.34$)  &$269.96$($\pm58.29$)  & $255.60$($\pm54.99$) &$268.16$($\pm54.79$) \\ 
 \cline{1-5}
 \end{tabular}}
 \caption{Means and standard deviations of the times participants spent playing the two games on the physical apparatus and in the {\em Levitation Simulator}.}
 \label{mean_playtime}
 \end{table}

\subsubsection{Observations and Interviews}
When we asked participants which game and in which setting they liked the most, there was no clear preference regarding the game. Most of the participants, however, preferred the interaction with the real prototype. 
Namely, nine participants preferred BeadBounce in the real condition, seven participants preferred LeviShooter in the real condition, and both games in the VR condition were preferred by four participants each.
Participants used words like \textit{new} and \textit{fun} to describe the game experience with the real prototype, and \textit{cool} and \textit{immersive} for the {\em Levitation Simulator}.
Almost all participants stated that the real and the virtual environment looked very similar, and in both cases they had a strong feeling of control.
Few participants noted that even though the experience was very similar, they had a 'different feeling' when playing the game in the real world and in VR - the interaction with the real prototype felt more intuitive to them.
Similarly as in the pointing user study, participants most frequently criticized the depth cues in VR and the resolution of the HMD, and reacted with more amazement when observing real levitation.

\section{Discussion}
Overall, the two user studies yielded similar results for the interaction in the {\em Levitation Simulator} and with the real prototype.
The overall results indicate that if a levitation user study is conducted in the {\em Levitation Simulator}, and then validated on the real prototype, there is a high probability that similar results will be obtained.

Certainly, differences might become significant with a larger sample size.
It is more important, however, to look at the effective sizes.
In the pointing study, for example, the average movement times and throughputs are relatively similar.
Thus we are quite confident, that any differences between the interfaces will not be very large in magnitude.

In both studies, we experienced two main limitations of the virtual environment.
These are important to keep in mind when developing or evaluating levitation applications in the {\em Levitation Simulator}. 

First, the quality of the depth cues with current VR headsets is still far from the depth cues one can perceive in a real user interface.
This effect is visible in the Fitts' law analysis and in the interview data of both user studies.
It might explain the trend in the pointing data, where smaller targets lead to a higher increase of movement times in the {\em Levitation Simulator} compared to the real condition (Fig.~\ref{fig:fittsVrLv}).
In the User Engagement study, this phenomenon was more prominent in the BeadBounce game, where the levitating particle and the head of the racket need to be at the same depth, for a hit to be registered.
We also observed that in the {\em Levitation Simulator} the increase in the movement time was steeper in the left-right direction, compared to front-back (Fig.~\ref{fig:fittsConditions}).
One possible explanation is that because of the lacking depth cues in VR, participants relied less on visual, and more on proprioceptive cues to complete the task. 
After identifying the correct target depth, in subsequent trials, they tried to reproduce the movement that resulted in a target hit the first time. 
From biomechanics perspective, however, it is easier to reproduce the front-back, then the left-right movement, since there are fewer joints involved (i.e. wrist in front-back, wrist and elbow or shoulder  in left-right). 

There are three ways to deal with the depth perception problem.
First, for tasks that require accurate depth perception, it is important to realize that performance in the {\em Levitation Simulator} might underestimate the performance that might be achieved with a real prototype.
Second, when evaluating virtual prototypes, it might be beneficial to choose tasks which do not rely as much on accurate depth perception.
This can be done by using larger targets overall, or by using targets which are larger in the depth dimension.
Third, we might hope that future VR headsets will provide improved depth perception.

The second shortcoming of the virtual prototype is that the ``wow effect'' seems to be gone.
The interview results and study observations indicate that the thrilling experience of interacting with a levitation interface, for the first time, cannot be replicated in VR.
This effect needs to be considered when estimating the user experience of a real levitation interface from a virtual prototype. 
The novelty appeal of physical levitation might partially explain the result that participants found the experience of playing the BeadBounce in the real condition more rewarding, compared to the VR condition (Fig.~\ref{fig:UES_BeadBounce}). 
This result, however, was not obtained for the LeviShooter game.
One possible reason might be a greater sense of involvement, through direct interaction in BeadBounce. 
Participants described the experience of reaching into the levitation volume with the racket and interacting with the levitating particle as unique and exciting, which possibly made the overall experience more rewarding and worthwhile.

Even though, in the interviews, the majority of the participants reported that they preferred playing the levitation games in the real condition, on average, they spent more time playing in VR (mean playtime = $270$~s for BeadBounce, $268$~s for LeviShooter).
The immersion property of VR, as well as the blocking of distractions, might be a possible explanation for this result. 
Thus it should be taken into consideration that the {\em Levitation Simulator} might overestimate the gameplay times, that would be obtained with the real prototype.

However, overall, the participants considered the interaction with the {\em Levitation Simulator} as highly realistic when comparing it to the real prototype.
We believe that, when accounted for differences in depth perception, immersion and the ``wow effect'', a virtual prototype can provide a very good prediction of the interaction performance and experience that would be achieved with a real prototype.

\section{Conclusion}
In this paper, we have presented the {\em Levitation Simulator}. 
We derived a model of the movement of a levitated particle in a sound field from first principles.
This makes the virtual levitating particle behave seemingly identical to a real particle.
Our user studies show that the {\em Levitation Simulator} provides performance as well as levels of user engagement and user experience comparable to the real levitation apparatus.
Here we demonstrated how virtual prototyping can be helpful in the design of user interfaces.
With this, we hope to inspire future research in modelling, simulation, and virtual prototyping of UIs in HCI.

\section{Acknowledgments}
This research has received funding from the European Union's Horizon 2020 research and innovation programme under grant agreement \#737087 (Levitate).
\balance{}

\balance{}

\bibliographystyle{SIGCHI-Reference-Format}
\bibliography{sample}

\end{document}